\documentclass[%
 amsmath,amssymb,
 aps,
]{emulateapj}

\usepackage{dcolumn}
\usepackage{bm}
\usepackage{graphicx}
\usepackage{epsfig}
\usepackage{color}
\usepackage[normalem]{ulem}

\usepackage{url}
\usepackage{hyperref}

\DeclareGraphicsExtensions{.pdf,.png,.jpg,.eps}

\begin{document}

\preprint{APS/123-QED}

\title{Systematic errors in low-latency gravitational wave parameter estimation impact electromagnetic follow-up observations}

\author{Tyson B. Littenberg}
\affiliation{Center for Interdisciplinary Exploration and Research in Astrophysics  (CIERA) and Department of Physics and Astronomy, Northwestern University, 2145 Sheridan Road, Evanston, IL  60208, USA}
\author{Ben Farr}
\affiliation{Enrico Fermi Institute, Department of Physics, and Kavli Institute for Cosmological Physics, University of Chicago, Chicago, IL 60637, USA}
\author{Scott Coughlin}
\affiliation{Cardiff University, Cardiff CF24 3AA, UK}
\author{Vicky Kalogera}
\affiliation{Center for Interdisciplinary Exploration and Research in Astrophysics  (CIERA) and Department of Physics and Astronomy, Northwestern University, 2145 Sheridan Road, Evanston, IL  60208}

\date{\today}

\begin{abstract}
Among the most eagerly anticipated opportunities made possible by Advanced LIGO/Virgo are multimessenger observations of compact mergers.
Optical counterparts may be short-lived so rapid characterization of gravitational wave (GW) events is paramount for discovering electromagnetic signatures.
One way to meet the demand for rapid GW parameter estimation is to trade off accuracy for speed, using waveform models with simplified treatment of the compact objects' spin.
We report on the systematic errors in GW parameter estimation suffered when using different spin approximations to recover generic signals. 
Component mass measurements can be biased by $>5\sigma$ using simple-precession waveforms and in excess of $20\sigma$ when non-spinning templates are employed.
This suggests that electromagnetic observing campaigns should not take a strict approach to selecting which LIGO/Virgo candidates warrant follow-up observations based on low-latency mass estimates.
For sky localization, we find searched areas are up to a factor of ${\sim}$2 larger for non-spinning analyses, and are systematically larger for any of the simplified waveforms considered in our analysis. 
Distance biases for the non-precessing waveforms can be in excess of 100\% and are largest when the spin angular momenta are in the orbital plane of the binary.  
We confirm that spin-aligned waveforms should be used for low-latency parameter estimation at the minimum.  Including simple precession, though more computationally costly, mitigates biases except for signals with extreme precession effects.
Our results shine a spotlight on the critical need for development of computationally inexpensive precessing waveforms and/or massively parallel algorithms for parameter estimation.  
\end{abstract}

\pacs{Valid PACS appear here}
\maketitle


\section{\label{sec:intro}Introduction}
Ground-based gravitational wave (GW) interferometers are most sensitive at frequencies between 10 Hz and 2 kHz.  
Binary systems comprising of compact stellar remnants such as black holes (BHs) and neutron stars (NSs) will experience the late stages of orbit inspiral and will merge at these frequencies, making them the prime target for the Advanced LIGO~\citep{aLIGO} and Advanced Virgo~\citep{aVirgo} network of detectors.
The scientific payoff from observing a population of compact binaries in GWs is hard to overstate.  
Mergers of binary neutron star (BNS) and/or neutron star-black hole (NSBH) systems are the leading candidates for the progenitors of short-hard gamma-ray bursts (sGRBs).
Comparing rates of GW detections and sGRBs will constrain the beaming angle for the gamma-ray jet while GW signatures from such systems coincident with GRBs will confirm the binary merger scenario~\citep{Chen:2013}.
Details of the inspiral and merger signals will probe regimes of strong-field, dynamical gravity inaccessible to any other observational or experimental techniques, testing the general relativistic field equations ~\citep{Cornish:2011ys, Li:2011cg, Sampson:2013lpa, Berti:2015itd} and examining the equation of state of matter at supranuclear density~\citep{Flanagan:2007ix, Read:2009yp, Lackey:2013axa, DelPozzo:2013ala, Wade:2014vqa}.
The merger rates and physical properties of compact binaries are sensitive to physical processes of compact object formation and binary evolution; their measurements will provide strong constraints on theoretical models for population synthesis~\citep{Belczynski:2009xy,Belczynski:2011bn,Mandel:2015spa}.  
Last but not least, well localized binaries will serve as cosmological ``standard sirens,'' further constraining the scale and expansion rate of the local universe~\citep{Schutz:1986gp,Holz:2005df,Nissanke:2009kt}.  

Clearly the advent of GW astronomy and astrophysics is highly anticipated in its own right. 
Furthermore, GW discoveries with identified electromagnetic (EM) counterparts would allow us to reach the full potential of multimessenger astronomy.
GW observatories are all-sky monitors and are well approximated by a quadrupole antenna with maximum sensitivity directly above and below the plane of the detector.
Because the antenna pattern varies slowly with angular distance away from the detector plane zenith, single detectors have poor sky-localization capabilities.
Differences in the arrival time, phase, and polarization of the GW signal in different detectors separated by large baselines enable more precise sky-location measurements because a relatively small region of the sky could produce a particular coherent signal.

Because the GW measurement will not be able to pinpoint the location of a source alone, great effort is needed to conduct electromagnetic observing campaigns.
Because any optical counterpart may be short-lived, and a typical EM telescope will have to use several pointings to cover the LIGO/Virgo error region on the sky, time is of the essence.
Unfortunately, obtaining the most precise and accurate estimates for a GW event's physical parameters is a time-consuming computational challenge.  
LIGO/Virgo parameter estimation is performed using stochastic Bayesian sampling algorithms such as Markov Chain Monte Carlo (MCMC)~\citep{Gamerman:1997}, Nested Sampling~\citep{Skilling}, or MultiNest~\citep{Feroz:2007kg}.  
Circularized binary inspiral signals are characterized by 15 parameters.
The 15-dimensional posterior distribution function is highly structured, often exhibiting multiple modes and strong correlations between parameters. 
The sampling algorithms perform a fundamentally serial random exploration of parameter space, at each point evaluating how well a trial signal matches the data.  
Each step of the analysis requires the computation of a model GW waveform, or ``template,'' which must be repeated $\mathcal{O}(10^7)$ times to sufficiently characterize the 15-dimensional parameter space.
The standard sampling algorithms take $\mathcal{O}({\rm days})$ of modern high-performance computing time to converge--latencies which can be longer in duration than the lifetime of proposed EM signatures such as kilonova~\citep{Metzger:2010}.

The two philosophical approaches to reducing the real-time latency for parameter estimation are to use computationally inexpensive approximations to the template models or a simplification of the likelihood function.
Several groups have explored novel, massively parallel, parameter estimation strategies to further reduce the time required for low-latency parameter estimation~\citep{Pankow:2015cra, Haster:2015, Farr:2015}.
While the massively parallel parameter estimation methods have shown promise in providing prompt, robust, signal characterization none have reached sufficient maturity to be deployed in the real-time analysis for the early Advanced LIGO observing runs.

GW parameters are often divided into \emph{extrinsic} parameters (orientation, sky location, distance), which depend on the observer, and \emph{intrinsic} parameters (masses and spins of the compact objects), which depend on the binary system. 
Generally speaking, intrinsic parameters influence the phase evolution of the binary waveform while extrinsic parameters are responsible for the relative amplitude, polarization, and time of arrival of the signal at each detector.
To meet the low-latency demands placed on GW parameter estimation for prompt EM follow-up, a hierarchical strategy has been devised that takes advantage of the separability between intrinsic and extrinsic parameters to rapidly report parameters for a candidate GW event (see~\cite{Farr:2014} for discussion on separate intrinsic and extrinsic analyses). 
In this paradigm computation time is saved by reducing the number of intrinsic degrees of freedom in the GW model by making simplifying assumptions about the spin of the compact objects, either assuming particular alignments for the spin angular momenta or fixing these terms to be zero.
The scheme first relies on the relative signal-to-noise ratio (S/N) and phase in each detector with latency of minutes~\citep{Singer:2015ema}, a parameter estimation analysis where the templates are restricted by assuming either non-spinning or aligned spin binaries (substantially decreasing run time to a few hours)~\citep{Singer:2014qca,Berry:2014jja}; finally the full parameter estimation analysis that uses generically spinning waveforms~\citep{Farr:2015lna}.

The separability of intrinsic and extrinsic parameters is a fair approximation when dealing with binary NS systems where spins are expected to be small, possibly less than 10\% of the maximum value allowed by general relativity.  This separability breaks down when the binary contains one or two BHs, as their spins can be near maximal (see ~\cite{Fragos:2015} and references therein).   Through relativistic couplings between the spin-spin and spin-orbit angular momenta, rapidly spinning misaligned binaries undergo dramatic precession of the orbital plane, thereby coupling the intrinsic and extrinsic parameters.  
As a result, restricting the spin degrees of freedom for template waveforms can bias the parameters we wish to measure rapidly,  namely sky location of the GW source and the distance to a binary signal, as well as to whether
the source involves a NS or not (the latter has implications for the potential existence of an EM counterpart).

\cite{Miller:2015sga} systematically compared the results of parameter estimation from using different waveforms, varying different spin parameterizations as well as post-Newtonian families, finding that biases in intrinsic parameter estimation (i.e. masses and spins) are comparable to the systematic differences between post-Newtonian families; they did not address issues related to extrinsic parameters of importance, such as sky location and distance.

In this paper, using a consistent post-Newtonian treatment for all waveforms, we put the biases due to reducing the spin degrees of freedom into astrophysical context. We address the following key question: do the spin-constrained, fast, frequency-domain waveforms planned for low-latency analysis introduce important biases in parameter estimation, specifically relevant to EM follow-up work (which is the reason for low-latency analysis)? 
We find that systematic errors incurred by using non-spinning or spin-aligned templates will be significant, with respect to both the statistical errors and the astrophysical interpretation of the results.  
For component masses that may be used to determine which candidate events should be followed up with telescopes, the biases can be in excess of several ${\rm M}_\odot$, opening  the possibility of misidentifying a NS as a BH or vice versa.
Electromagnetic searches will have to cover an area on the sky larger by a factor of $\lesssim 2$ to have the same probability of pointing at the true source location compared to the sky localization of the fully spinning/precessing analysis.  
While the biases systematically decrease as we use increasingly complete waveforms, we find no substitute, fast, frequency-domain template waveform that is immune to introducing substantial biases in some parts of parameter space.

In the near term our results should be used to guide follow-up observing strategies.   For the future we hope that our finding helps stimulate further development of fast, accurate waveforms that contain all of the necessary degrees of freedom for accurate parameter estimation.

\section{\label{sec:setup} Setup}

We simulate 1000 LIGO/Virgo signals randomly selected from a uniform distribution in component mass ($m_1,m_2$) drawn from $m\in[1,30]$ ${\rm M}_\odot$ with an additional constraint on the total mass of the binary $m_1+m_2\leq30$ ${\rm M}_\odot$; six parameters describing the two spin vectors $\boldsymbol{S}_1,\boldsymbol{S}_2$ with magnitudes drawn from $U[0,1]$ and isotropic orientation with respect to the orbital angular momentum $\boldsymbol{L}$; sky location ($\alpha,\delta$) distributed uniformly on the celestial sphere; and three angles drawn uniformly over a sphere that describe the orientation of the binary ($\theta_{J,N}, \psi, \phi_0$).  The $\theta_{J,N}$ parameter is the angle between the wave propagation direction $\hat N$ (opposite the observer's line of sight $\boldsymbol k$) and the total angular momentum $\boldsymbol J$.  In the limit of non-spinning binaries this is degenerate with the inclination angle $\iota=\cos^{-1}(\hat k \cdot \hat L)$ which is more traditionally used in the GW literature.   We prefer $\theta_{J,N}$ because it has less variation during the orbital evolution, while the direction of $\boldsymbol L$ is time-varying for spin configurations that induce significant precession~\citep{Farr:2014qka}.  
The luminosity distance, $D_L$, of each source is drawn uniformly in volume with an additional detectability requirement that the binary has a signal-to-noise ratio ${\rm S/N}>5$ in at least two detectors.
This S/N requirement produces a strong selection effect due to the quadrupolar emission of gravitational radiation.  
While GW emission from a compact binary is omnidirectional, it is strongest along the direction of the orbital angular momentum vector $\mathbf L$ of the binary.
A system that is ``face on,'' i.e. the line of sight to the observer $\mathbf k$ and $\mathbf L$, is nearly parallel ($\hat k \cdot \hat L \sim 1$), is detectable out to a larger distance than the same system in an ``edge on'' ( $\hat k \cdot \hat L \sim 0$ ) configuration.
Assuming that compact binaries are distributed uniformly in volume, far more face-on binaries will be detected than edge-on binaries (see Fig. 2 and related discussion in \cite{Littenberg:2015tpa}.

In our simulations, we generate the source GW signals using the SpinTaylorT2 waveforms, which are time-domain inspiral-only waveforms that allow for generic spin magnitude and orientation, and hence for full precession.  The waveforms do not include any treatment of the merger or ringdown signal, and assume quasi-circular orbits and negligible finite-size effects of the components.
For each binary we compute the response of Advanced LIGO/Virgo at design sensitivity~\cite{ObservingScenarios} with a low-frequency cutoff at 20 Hz. Each simulated signal is processed by the LIGO/Virgo data analysis pipeline for parameter estimation, LALInference.  
LALInference is a code library built to provide robust Bayesian analysis of GW signals and is part of the LIGO Analysis Library ~\citep{LAL}.
For this study we use the \textsc{lalinference\_mcmc} application in LALInference, a parallel tempered MCMC algorithm.
A full description of the LALInference package, and the custom features found in lalinference\_mcmc for analysis of compact binaries, can be found in \cite{LALInference}.  

\begin{figure}
   \centering
   \includegraphics[width=\linewidth]{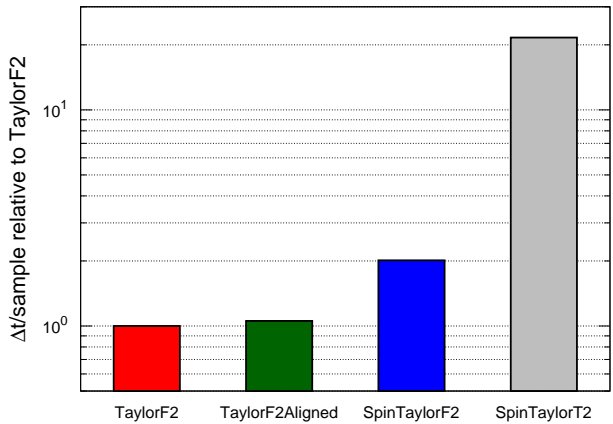} 
   \caption{\small{Relative speed for different waveforms tested in this work estimated by drawing 1000 independent samples from the lalinference\_mcmc prior.  All times are measured relative to the TaylorF2 (non-spinning) waveforms.  Spin-aligned (TaylorF2Aligned) waveforms add negligible cost.  Simple precession (SpinTaylorF2) is a factor of ${\sim}2$ more costly, while the full precession waveform (SpinTaylorT2) is slower by an order of magnitude.  These results do not take into account the added inefficiency of sampling the higher dimension posteriors. }}
   \label{fig:benchmark}
\end{figure}

The LALInference analysis is repeated using four template waveforms:  SpinTaylorT2, which provides bias-free parameter estimation of our simulated signals (``full precession''), and three other faster, frequency-domain waveforms, all inspiral-only, for consistency with the source signals: SpinTaylorF2, which uses a simplified treatment of the precession that conserves the opening angle $\beta = \cos^{-1}\left( \hat{J}\cdot \hat{L}\right)$ (``simple precession''); TaylorF2Aligned, which fits for spin magnitudes assuming $\beta=0$ (``aligned spin''); and zero-spin TaylorF2 waveforms (``non-spinning'').  
Figure~\ref{fig:benchmark} shows the computational cost per likelihood evaluation between the different waveforms employed in this work.  The approximate cost for each waveform model is determined by using the lalinference\_mcmc software to draw 1000 independent samples for the prior distribution.  The results are normalized by the computation time for the non-spinning TaylorF2 waveforms because they are the fastest available and the baseline model for rapid parameter estimation.  Including the two spin magnitudes in a constrained, fully aligned configuration (TaylorF2aligned) has a negligible effect on the computational cost of the model.  Allowing for simple precession (SpinTaylorF2) comes with a factor of ${\sim}2$ increase in run time, while the time-domain, fully precessing waveform (SpinTaylorT2) is over an order of magnitude more computationally taxing.  This simple benchmarking test does not take into account the increased difficulty of sampling higher dimension posteriors, which can increase the convergence time for the MCMC analysis, but serves as a useful reference for the tradeoff between waveform accuracy and speed.

The output of lalinference\_mcmc is a series of independent samples drawn from the posterior distribution function for the template parameters that are used to draw inferences about the astrophysical nature of the GW source.  

\section{\label{sec:int} Results: Intrinsic parameters}

We begin by focusing on intrinsic parameter biases.
Because the spin and mass parameters govern the phase evolution of the binary, artificially constraining spin degrees of freedom for generic binaries will have a dramatic effect on mass measurement.
\begin{figure*}
   \centering
   \includegraphics[width=\linewidth]{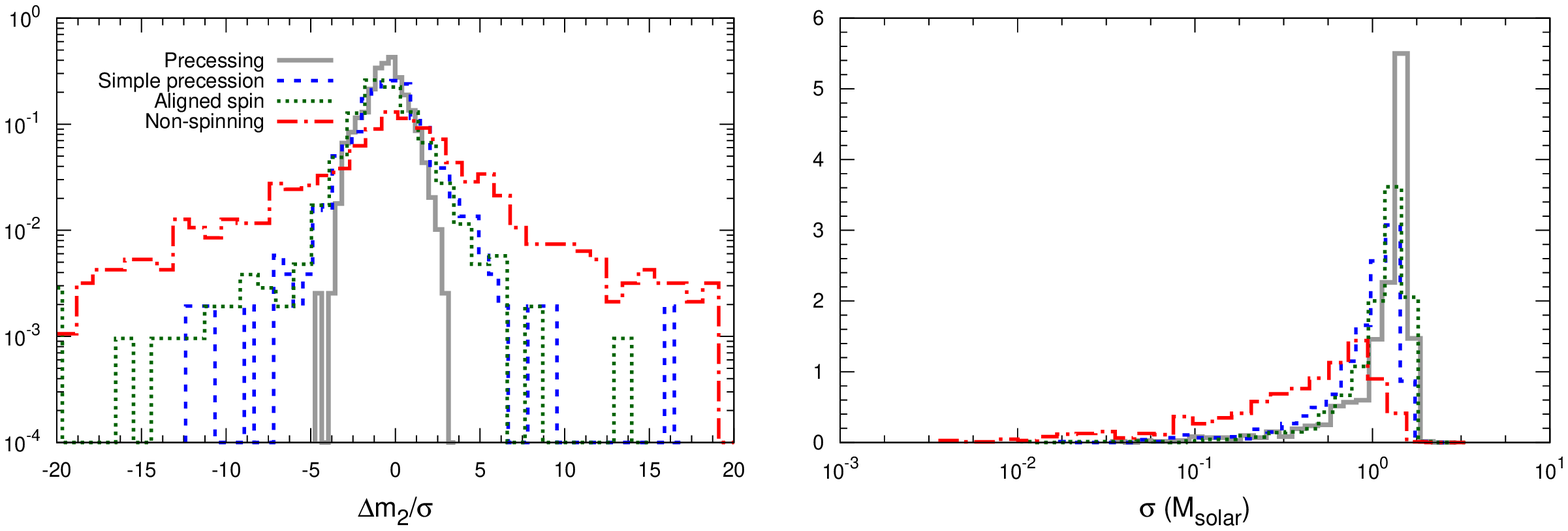} 
   \caption{\small{Left: bias in component mass recovery for full precession (gray, solid), simple precession (blue, dashed), aligned spin (green, dotted), and non-spinning (red, dotted-dashed) templates measured in standard deviations of the full-precession posterior.  We find large biases when using simplified treatments of the spin parameters.  Right: same color scheme as the left panel, now showing the distribution of the standard deviation for the component mass posteriors, which are typically ${\sim}1$ ${\rm M}_\odot$}.  The biases we find are astrophysically significant.}
   \label{fig:massbias}
\end{figure*}
Figure~\ref{fig:massbias} summarizes the consequences of restricting spin degrees of freedom for mass determination.
Results shown in this figure come from marginalized posterior distributions for the smaller mass in the binary, $m_2$.  We use $m_2$ as a proxy for whether a GW signal is a good candidate for electromagnetic observations.  While BNS and NSBH systems are considered possible progenitors of short-duration GRBs, there are no strong theoretical candidates from electromagnetic signals from merging stellar-mass BHs.  A resource-limited electromagnetic follow-up campaign may consider using the inferred $m_2$ to determine which candidate LIGO/Virgo detections are worth telescope time, selecting only those systems that have an $m_2$ measurement consistent with NS masses.

The left panel is a histogram of the bias on $m_2$.  The bias is measured as the difference $\Delta\,m_2$ between the mean of the $m_2$ posterior and the true value, normalized by the posterior's standard deviation $\sigma$.  Using this definition, the bias is a measure of how many standard deviations ($\sigma$) away from the true value the posterior distribution is peaked.  In this figure and throughout this paper, the gray, blue, green, and red curves correspond to the precessing, simple precession, aligned spin, and non-spinning waveforms, respectively.  In Fig.~\ref{fig:massbias}  we find that the possible biases can be in excess of $5\sigma$ even when using the simple-precession waveforms, and reach beyond $20\sigma$ for the non-spinning waveforms often used for low-latency parameter estimation.  The right panel is the distribution of standard deviations for the same simulated signals, which are strongly peaked around 1 ${\rm M}_\odot$ with a long tail down to ${\sim}10^{-2}$  ${\rm M}_\odot$ for the non-spinning waveforms.  Not only is a $10\sigma$ or $20\sigma$ bias statistically alarming, but because of the typical standard deviations for this parameter, these biases can have astrophysical repercussions as well~\citep{Littenberg:2015tpa}. 
\begin{figure*}
   \centering
   \includegraphics[width=\linewidth]{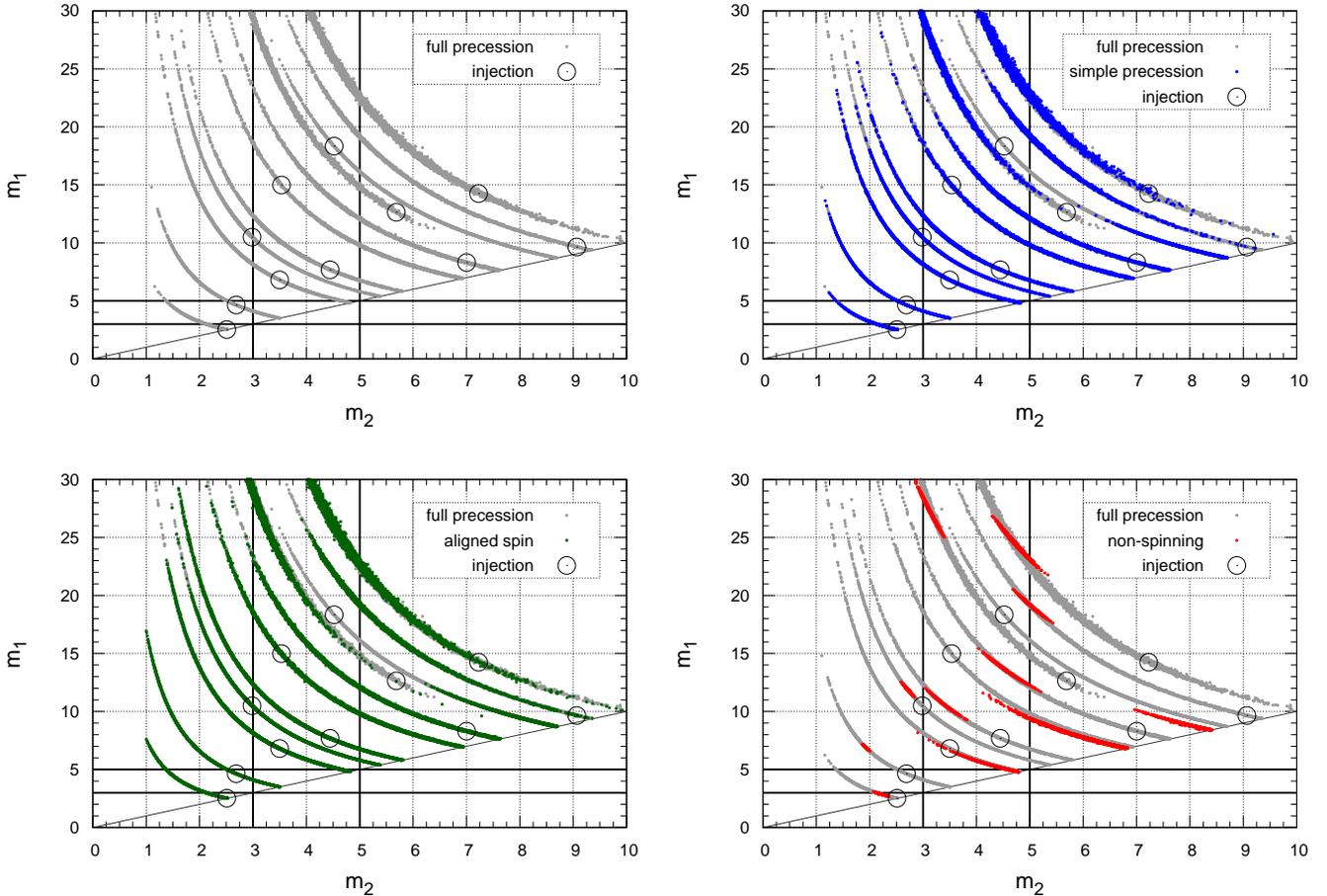} 
   \caption{\small{A selection of $m_1$--$m_2$ posteriors compared to the true source values (open circles) for the four different spin configurations:  full precession (gray points in each panel, plotted by themselves in the top left; simple precession (blue, top right); aligned spin (green, bottom left); and non-spinning (red, bottom right).   Heavy vertical and horizontal lines mark the ``mass gap'' boundary between high-mass NSs and low-mass BHs.  The inferred component masses from non-spinning waveforms generically miss the true source value and often incorrectly classify the binary.  Including aligned spins or simple precession improves the agreement with the full-precession results but misclassification is still possible.  Low-latency mass estimates should be taken with caution.}}
   \label{fig:mass}
\end{figure*}

Figure~\ref{fig:mass} shows 11 randomly selected binaries (open circles) from our simulated population with a scatter plot of the $m_1$--$m_2$ posterior samples for each source.  The heavy vertical and horizontal lines denote the putative ``mass gap'' between the theorized highest mass NSs (${\sim}3$ M$_\odot$) and the observed lowest mass BHs (${\sim}5$ M$_\odot$)~\citep{Farr:2010tu, Ozel:2010su, Kreidberg:2012ud}.  Each panel shows the recovered masses for the different waveforms tested superimposed with the bias-free full precession posteriors, which are shown in gray in each panel, and alone in the top left.  Here we find that the non-spinning waveforms (red, bottom right) are highly prone to misclassifying the GW source.  The true value (open circle) is typically well outside of the region in $m_1$--$m_2$ space covered by the posterior and often confined in the wrong region of parameter space entirely (e.g. in the NSBH region when the true signal was a binary BH of comparable mass).  When including some freedom in the spin parameters, whether that be the aligned (green, bottom left) or simple precession case (blue, top right), we find more consistent overlap with the bias-free posteriors.  However, neither the aligned spin nor the simple-precession waveforms can generically be substituted for the full precession waveforms without risking some substantial biases in the determination of $m_1$ and $m_2$.   For example, both the aligned spin and simple-precession waveforms misidentified a source with a low-mass secondary ($m_1\sim19$M$_\odot$ and  $m_2\sim4.5$M$_\odot$) as a binary BH system of comparable mass.  
We conclude that low-latency mass estimates are not generically reliable, and that electromagnetic observing campaigns should not take a strict approach to selecting which LIGO/Virgo candidates warrant follow-up observations based on early mass estimates. 

\section{\label{sec:ext} Results: Extrinsic parameters}
The most pertinent measurements needed for low-latency GW searches are the sky location and distance to the source.  
The sky location is of obvious importance for follow-up electromagnetic observations, and the distance can be used to either select nearby candidate GW sources for observation or to convolve the GW measurement with galaxy catalogs~\citep{Nissanke:2012dj,Hanna:2013yda,Singer:2014qca, Fan:2014kka}.

For sky localization, our figure of merit is the ``searched area'' for different waveforms.
The searched area is a measure of the number of square degrees contained within the credible interval of the posterior that intersects the true signal parameters.  If an electromagnetic follow-up strategy is to sequentially point a telescope at fields on the sky ranked by their posterior weight, the searched area gives a sense of how large a region needs to be searched to have a certain probability of the true source location being included among the observations.

\begin{figure}
   \centering
   \includegraphics[width=\linewidth]{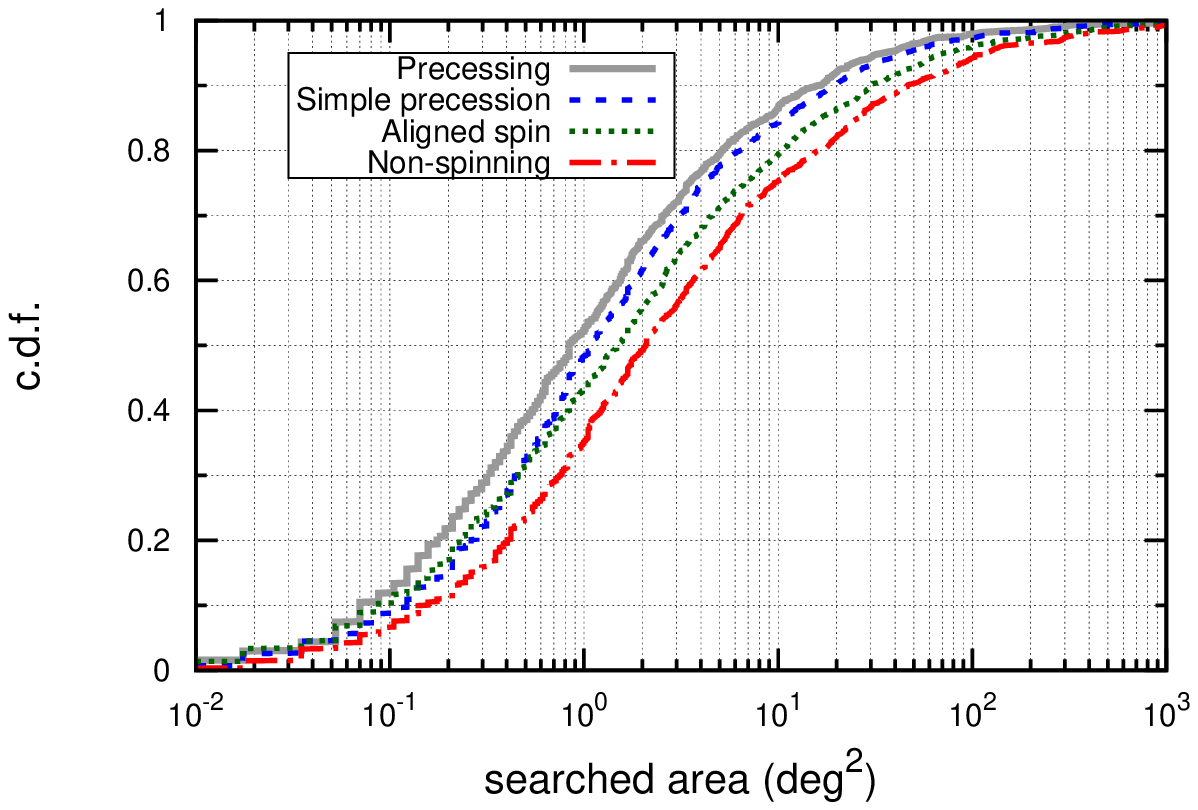} 
   \caption{\small{Cumulative distribution functions for the searched sky area of each injection when using the non-spinning templates (red, dotted-dashed), aligned spin (green, dotted), simple precession (blue, dashed), and the precessing templates (gray, solid).  The non-spinning analysis systematically requires more searched area to find the true signal.  Including more accuracy in the waveform systematically improves searched areas.  None of the simplified waveforms matches the performance of the fully precessing templates.  }}
   \label{fig:area}
\end{figure}

Figure~\ref{fig:area} shows the cumulative distribution function of searched area for the precessing, simple precession, aligned spin, and non-spinning waveforms.  We see that searched areas are systematically shifted to larger values as the spin degrees of freedom are further restricted, meaning that sky-location biases incurred by the simplified waveforms translate to more telescope pointings to ensure the same probability of imaging the true source.  For most signals in our simulated population, the differences in searched area between the fully precessing waveforms and the non-spinning waveforms are around a factor of two.  Using spin-aligned waveforms yields searched areas that are typically ${\sim}30\%$ to ${\sim}50\%$ larger than the full-precession results. Simple-precession searched areas are, at most, ${\sim}30\%$ larger but typically are within ${\sim}20\%$ of the full-precession areas.

The distance to the source is a more difficult parameter to measure.  As noted throughout the GW literature, first in \cite{Cutler:1994ys} and further investigated in \cite{Nissanke:2009kt} and \cite{Rodriguez:2013oaa}, there is a strong degeneracy between the orientation of the binary and the distance which causes the waveforms of near by, edge-on systems to be nearly indistinguishable from distant, face-on systems at plausible S/Ns for Advanced LIGO and Virgo observations.  Convolving this with the assumption that binaries are uniformly distributed within the LIGO/Virgo horizon results in distance posteriors heavily influenced by the prior, which strongly favors distant, face-on systems.

\begin{figure*}
   \centering
   \includegraphics[width=\linewidth]{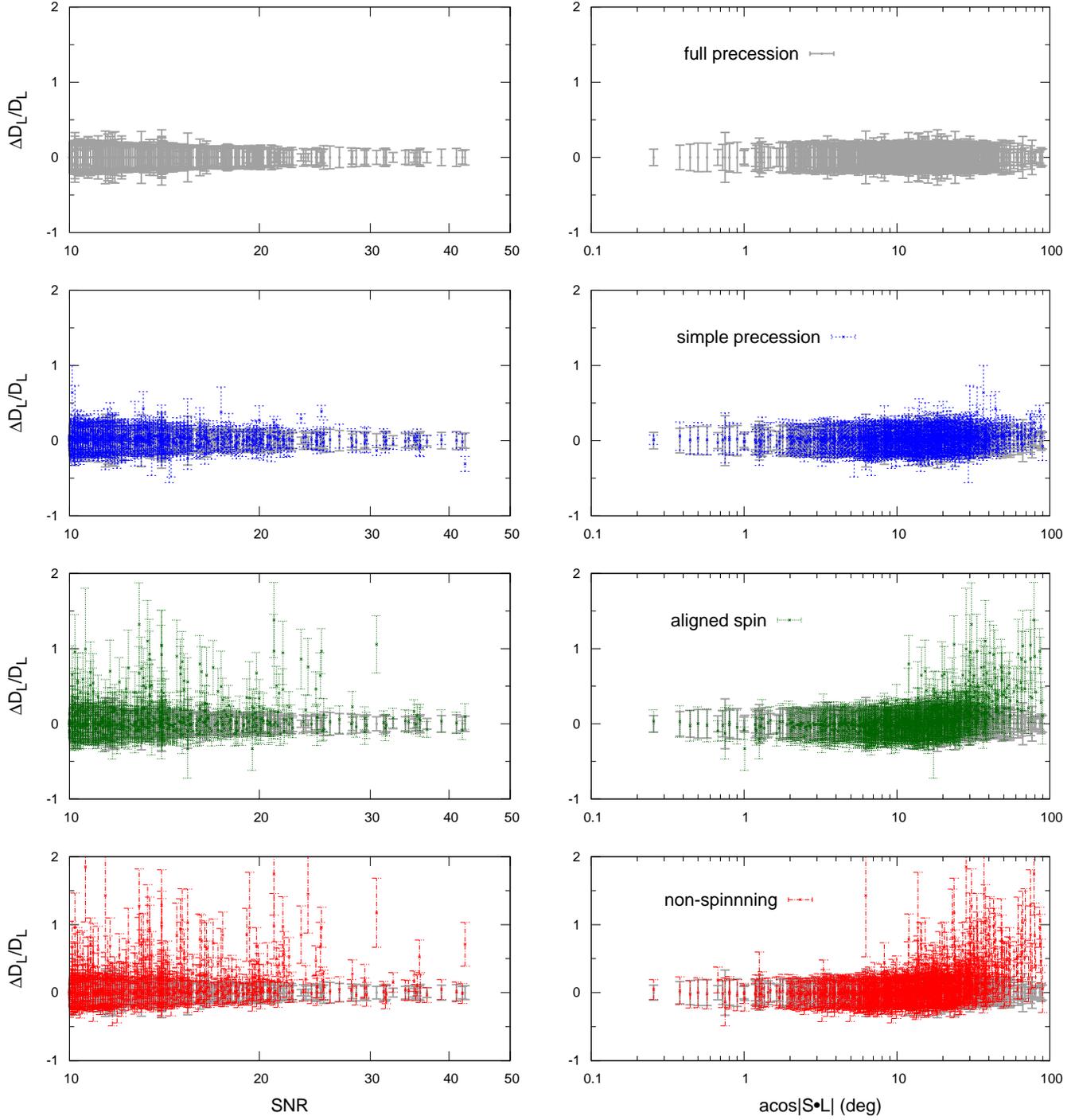} 
   \caption{\small{Left column:  fractional differences in distance as a function of S/N.  From top to bottom the results correspond to the full precession, simple precession, aligned spin, and non-spinning analyses.  Systematic errors in distance are largest for waveforms without precession and the errors do not strongly scale with S/N.  Right column:  same ordering but now the independent variable is the opening angle $\beta$ between the total and orbital angular momenta.  Here we see that the bias is strongest for opening angles close to $90^\circ$, which are cases that exhibit the most dramatic precession effects.}}
   \label{fig:dLbiasAll}
\end{figure*}

To quantify the difference in distance measurement we use the bias-free fully precessing distance estimates as the baseline against which the other waveforms' performances are measured.
Figure~\ref{fig:dLbiasAll} shows the fractional difference $\Delta D_L/D_L$ in distance $D_L$, defined as the difference between the peaks of the distance posteriors for the waveform under investigation and the full-precession waveform divided by the peak of the full-precession distance posterior.  The error bars in Figure~\ref{fig:dLbiasAll} are one standard deviation of the posterior normalized by the peak of the full-precession distance distribution.  The result is a measure of the bias and fractional error in distance measurement from using waveforms that make simplifying approximations to the spin.  From top to bottom, the rows of plots correspond to the full precession (gray), simple precession (blue), aligned spin (green), and non-spinning (red) templates used for recovery.  Our use of the mode of the distance posterior as the point estimate against which we make comparisons was for convenience.  Macroscopic differences between the distributions are insensitive to the particular choice of definition for bias in distance.   

The left column in Figure~\ref{fig:dLbiasAll} shows $\frac{\Delta D_L}{D_L}$ as a function of the true signal's network S/N.  
When using the same waveform for recovery as we do for simulating the data (top row) we find statistical errors of 10-30\%.  We find that the distance biases grow and exceed the statistical uncertainty as the waveforms employed for recovery use simpler treatment of the spin.  The biases in distance determination are not found to be strongly dependent on S/N.  As a consequence, for accurate distance estimation spin precession will have to be considered for any detectable signal.

The right column in Figure~\ref{fig:dLbiasAll} presents the fractional error in distance $\frac{\Delta D_L}{D_L}$ as a function of the opening angle $/beta$ between total and orbital angular momenta.
Here we see that the biases become larger than the statistical error for the aligned (third row, green) and non-spinning (bottom row, red) waveforms for $\beta>10^\circ$ and can be larger than 100\% for $\beta>30^\circ$.  At $\beta=90^\circ$ the spin angular momentum vector lies in the orbital plane of the binary, maximizing the orbital precession.  The aligned-spin and non-spinning waveforms assume a constant binary orientation, making these waveforms a poor match to the true signal.  Except for rare instances the simple-precession waveforms (second row, blue) return qualitatively similar distance posteriors to the full-precession waveforms.

None of the results shown have assumed anything about the GW source beyond the implicit validity of the waveform models used for signal recovery.
A special case for ground-based GW astrophysics is the simultaneous observation of the GW and EM signature from a compact merger.  
The most plausible EM counterparts to binary mergers involving at least one NS are short GRBs.  It is generally assumed that the gamma-rays are emitted along a narrow, relativistic jet with an opening angle of order $10^\circ$.  In a joint GRB/GW analysis, assuming the jet is parallel to the angular momentum vector, gamma ray detections place a strong constraint on the allowed values of $\theta_{J,N}$, and consequently $D_L$ for the binary.  

\begin{figure}
   \centering
   \includegraphics[width=\linewidth]{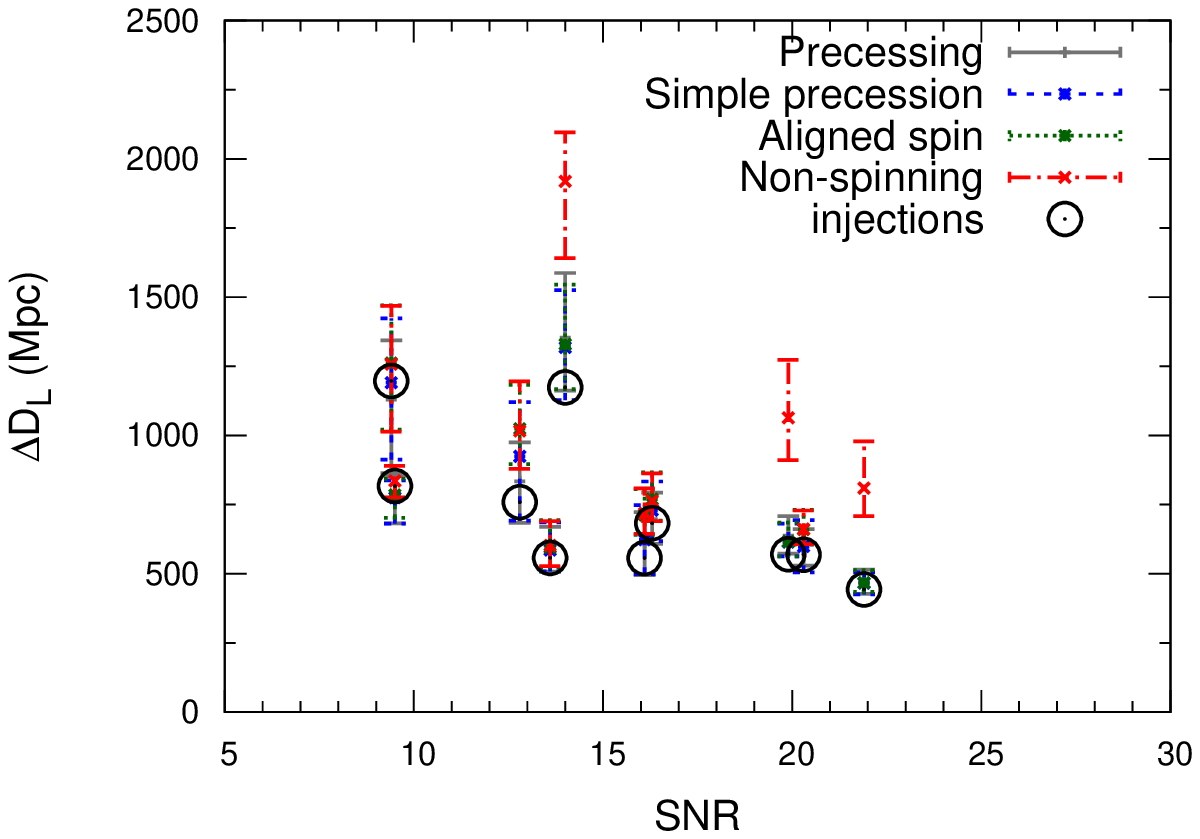} 
   \caption{\small{Examples of marginalized distance posteriors for full precession (gray, solid), simple precession (blue, dashed), aligned spin (green, dotted), and non-spinning (red, dotted-dashed) distance posteriors.  Results correspond to the NSBH systems in our injection set with face-on ($\theta_{J,N}<20^\circ$) orientations.  Black circles mark the true value.  Points are the median values and error bars span the 90\% credible interval of the posterior. For all cases the simple precession and precessing recovery include the injection.  The aligned-spin analysis excludes the true distance for three of the 10 simulations.  The non-spinning analysis excludes the true distance in all but four of the simulations.}}
   \label{fig:dLbiasGRB}
\end{figure}

For a sense of how the inferred distance depends on different spin approximations, we select NSBH events from our population of simulated signals that had $\theta_{J,N} < 20^\circ$ (or $>160^\circ$) as a set of potential short GRBs.  If coincident GRB detections were made we could adopt a strict prior on the orientation of the binary, which would help constrain the distance to the source. 
Figure~\ref{fig:dLbiasGRB} shows $90\%$ credible intervals for the distance when we adopt the ``GRB prior'' for $\theta_{J,N}$ as a function of the true signals' S/N.  
The black circles denote the true distance.  Once more we find significant bias in the distance estimate when non-spinning (red, dotted-dashed) waveforms are used for the analysis. The aligned spin (green, dotted) analysis excludes the true distance for three of the 10 simulation.  For this small sample the simple precession (blue, dashed) and full precession (gray, solid) waveforms return qualitatively similar distance posteriors, and in only one event do they differ significantly.

\section{\label{sec:discuss} Discussion}

Among the most eagerly anticipated discoveries made possible by Advanced LIGO/Virgo are joint GW and EM observations of compact mergers.
Optical counterparts of compact mergers may be short-lived so a rapid response of electromagnetic follow-up observations to GW triggers is paramount for discovering the EM signature of GW detections.
One way to meet the demand for rapid GW parameter estimation is to trade off accuracy for speed, using computationally inexpensive waveform models that make simplifying assumptions about the spin of the constituent compact objects.

In this paper we studied the biases in parameter estimation for generic GW signals when non-spinning, aligned-spin, and simple-precession templates are used for the analysis.  We find non-negligible systematic errors on both intrinsic (masses) and extrinsic (location and orientation) parameters.  
Our study builds on \cite{Miller:2015sga}, which investigated systematic differences between post-Newtonian waveforms, by putting the biases resulting from restricting the spin degrees of freedom into astrophysical context.   We conclude that the biases in parameter estimation of spinning binaries using non-spinning templates will be large enough to impact astrophysical interpretation of the results.  

Component mass measurements, which may be used to determine which candidate events should be followed up with telescopes, can be biased by $>5\sigma$ using simple-precession waveforms and in excess of $20\sigma$ when non-spinning templates are employed.  These systematic errors translate to biases several ${\rm M}_\odot$ large, potentially leading to misclassification of NSs and BHs.
This suggests that electromagnetic observing campaigns should not take a strict approach to selecting which LIGO/Virgo candidates warrant follow-up observations based on low-latency mass estimates.

Without question the most pertinent parameter estimation result of GW searches is the inferred sky location of the source.  We find that searched areas are up to a factor of ${\sim}$2 larger for non-spinning analyses, and are systematically larger for any of the simplified waveforms considered in our analysis.  
While the searched areas systematically decrease as we use increasingly complete waveforms, we find no substitute template waveform that is able to reproduce the sky-localization performance of the fully spinning/precessing waveforms.

Distance determination can be used for selecting which candidates to observe, optimizing exposure times, and employing galaxy catalogs to help guide EM observations.  
Distance biases for the non-precessing waveforms can be in excess of 100\% and are largest when the spin angular momenta are in the orbital plane of the binary.  
The simple-precession waveforms produce consistently similar distance posteriors to the full-precession waveforms and are the only reliable low-latency option if distance estimates are deemed important for electromagnetic follow-up observations.
For potential GRB sources in our simulated population--NSBH injections with orbital angular momentum oriented within $20^\circ$--the simple-precession waveforms produced distance estimates consistent with the full-precession waveforms while the aligned-spin recovery excluded the true distance at the 90\% level for three of the 10 injections.  The non-spinning waveforms showed a substantial bias, often overestimating the distance to the source by hundreds of Mpc.

In the near term, we confirm that spin-aligned waveforms should be used for low-latency parameter estimation at the minimum.  The computational cost is nearly identical to that of the non-spinning waveforms, and the inclusion of spin dramatically improves mass estimates and distance estimates for face-on systems.  
The aligned-spin models are still subject to significant biases in the distance when the spin angular momenta are in the orbital plane of the binary, and yield larger searched areas than the fully precessing waveforms. At the added cost of a factor of ${\sim}2$ in computational effort, the simple-precession waveforms correct biases in the distance and make further improvements to the sky localization.  However, at this time there is no substitute for the full spinning precessing analysis for accurate parameter estimation in any of the parameters we studied.
Our results shine a spotlight on the importance of efforts to develop computationally inexpensive precessing waveforms~\citep{Field:2013cfa,Lundgren:2013jla, Chatziioannou:2014bma,Klein:2014bua}, and for novel, embarrassingly parallel, parameter estimation methods \citep{Pankow:2015cra, Haster:2015, Farr:2015}.  There is a critical need for low-latency parameter estimation methods which are able to account for all known features in the signals.

\section{Acknowledgments}
We thank Atul Adhikari, Claudeson Azurin, Brian Klein, Brandon Miller, Leah Perri, Ben Sandeen, Jeremy Vollen, Michael Zevin for help running the MCMC analysis.  
T.B.L. and V.K. acknowledge NSF award PHY-1307020.  B.F. was supported by the Enrico Fermi Institute at the University of Chicago as a McCormick Fellow. S.C. thanks the US-UK Fulbright Commission for personal financial support during this research period.  
Computational resources were provided by the Northwestern University Grail cluster through NSF MRI award PHY-1126812.


\begin{thebibliography}{49}%
\makeatletter
\providecommand \@ifxundefined [1]{%
 \@ifx{#1\undefined}
}%
\providecommand \@ifnum [1]{%
 \ifnum #1\expandafter \@firstoftwo
 \else \expandafter \@secondoftwo
 \fi
}%
\providecommand \@ifx [1]{%
 \ifx #1\expandafter \@firstoftwo
 \else \expandafter \@secondoftwo
 \fi
}%
\providecommand \natexlab [1]{#1}%
\providecommand \enquote  [1]{``#1''}%
\providecommand \bibnamefont  [1]{#1}%
\providecommand \bibfnamefont [1]{#1}%
\providecommand \citenamefont [1]{#1}%
\providecommand \href@noop [0]{\@secondoftwo}%
\providecommand \href [0]{\begingroup \@sanitize@url \@href}%
\providecommand \@href[1]{\@@startlink{#1}\@@href}%
\providecommand \@@href[1]{\endgroup#1\@@endlink}%
\providecommand \@sanitize@url [0]{\catcode `\\12\catcode `\$12\catcode
  `\&12\catcode `\#12\catcode `\^12\catcode `\_12\catcode `\%12\relax}%
\providecommand \@@startlink[1]{}%
\providecommand \@@endlink[0]{}%
\providecommand \url  [0]{\begingroup\@sanitize@url \@url }%
\providecommand \@url [1]{\endgroup\@href {#1}{\urlprefix }}%
\providecommand \urlprefix  [0]{URL }%
\providecommand \Eprint [0]{\href }%
\providecommand \doibase [0]{http://dx.doi.org/}%
\providecommand \selectlanguage [0]{\@gobble}%
\providecommand \bibinfo  [0]{\@secondoftwo}%
\providecommand \bibfield  [0]{\@secondoftwo}%
\providecommand \translation [1]{[#1]}%
\providecommand \BibitemOpen [0]{}%
\providecommand \bibitemStop [0]{}%
\providecommand \bibitemNoStop [0]{.\EOS\space}%
\providecommand \EOS [0]{\spacefactor3000\relax}%
\providecommand \BibitemShut  [1]{\csname bibitem#1\endcsname}%
\let\auto@bib@innerbib\@empty
\bibitem [{\citenamefont {Abadie}\ \emph {et~al.}(2014)\citenamefont {Abadie}
  \emph {et~al.}}]{aLIGO}%
  \BibitemOpen
  \bibfield  {author} {\bibinfo {author} {\bibfnamefont {J.}~\bibnamefont
  {Abadie}} \emph {et~al.} (\bibinfo {collaboration} {The LIGO Scientific
  Collaboration}),\ }\href@noop {} {\  (\bibinfo {year} {2014})},\ \Eprint
  {http://arxiv.org/abs/1411.4547} {arXiv:1411.4547 [gr-qc]} \BibitemShut
  {NoStop}%
\bibitem [{\citenamefont {Acernese}\ \emph {et~al.}(2015)\citenamefont
  {Acernese} \emph {et~al.}}]{aVirgo}%
  \BibitemOpen
  \bibfield  {author} {\bibinfo {author} {\bibfnamefont {F.}~\bibnamefont
  {Acernese}} \emph {et~al.} (\bibinfo {collaboration} {Virgo Collaboration}),\
  }\href {\doibase 10.1088/0264-9381/32/2/024001} {\bibfield  {journal}
  {\bibinfo  {journal} {Class.Quant.Grav.}\ }\textbf {\bibinfo {volume} {32}},\
  \bibinfo {pages} {024001} (\bibinfo {year} {2015})},\ \Eprint
  {http://arxiv.org/abs/1408.3978} {arXiv:1408.3978 [gr-qc]} \BibitemShut
  {NoStop}%
\bibitem [{\citenamefont {{Chen}}\ and\ \citenamefont
  {{Holz}}(2013)}]{Chen:2013}%
  \BibitemOpen
  \bibfield  {author} {\bibinfo {author} {\bibfnamefont {H.-Y.}\ \bibnamefont
  {{Chen}}}\ and\ \bibinfo {author} {\bibfnamefont {D.~E.}\ \bibnamefont
  {{Holz}}},\ }\href {\doibase 10.1103/PhysRevLett.111.181101} {\bibfield
  {journal} {\bibinfo  {journal} {Physical Review Letters}\ }\textbf {\bibinfo
  {volume} {111}},\ \bibinfo {eid} {181101} (\bibinfo {year} {2013})},\ \Eprint
  {http://arxiv.org/abs/1206.0703} {arXiv:1206.0703} \BibitemShut {NoStop}%
\bibitem [{\citenamefont {Cornish}\ \emph {et~al.}(2011)\citenamefont
  {Cornish}, \citenamefont {Sampson}, \citenamefont {Yunes},\ and\
  \citenamefont {Pretorius}}]{Cornish:2011ys}%
  \BibitemOpen
  \bibfield  {author} {\bibinfo {author} {\bibfnamefont {N.}~\bibnamefont
  {Cornish}}, \bibinfo {author} {\bibfnamefont {L.}~\bibnamefont {Sampson}},
  \bibinfo {author} {\bibfnamefont {N.}~\bibnamefont {Yunes}}, \ and\ \bibinfo
  {author} {\bibfnamefont {F.}~\bibnamefont {Pretorius}},\ }\href {\doibase
  10.1103/PhysRevD.84.062003} {\bibfield  {journal} {\bibinfo  {journal} {Phys.
  Rev. D}\ }\textbf {\bibinfo {volume} {84}},\ \bibinfo {pages} {062003}
  (\bibinfo {year} {2011})},\ \Eprint {http://arxiv.org/abs/{arXiv:1105.2088
  [gr-qc]}} {{arXiv:1105.2088 [gr-qc]}} \BibitemShut {NoStop}%
\bibitem [{\citenamefont {Li}\ \emph {et~al.}(2012)\citenamefont {Li},
  \citenamefont {Del~Pozzo}, \citenamefont {Vitale}, \citenamefont {Van
  Den~Broeck}, \citenamefont {Agathos} \emph {et~al.}}]{Li:2011cg}%
  \BibitemOpen
  \bibfield  {author} {\bibinfo {author} {\bibfnamefont {T.}~\bibnamefont
  {Li}}, \bibinfo {author} {\bibfnamefont {W.}~\bibnamefont {Del~Pozzo}},
  \bibinfo {author} {\bibfnamefont {S.}~\bibnamefont {Vitale}}, \bibinfo
  {author} {\bibfnamefont {C.}~\bibnamefont {Van Den~Broeck}}, \bibinfo
  {author} {\bibfnamefont {M.}~\bibnamefont {Agathos}},  \emph {et~al.},\
  }\href {\doibase 10.1103/PhysRevD.85.082003} {\bibfield  {journal} {\bibinfo
  {journal} {Phys.Rev.}\ }\textbf {\bibinfo {volume} {D85}},\ \bibinfo {pages}
  {082003} (\bibinfo {year} {2012})},\ \Eprint {http://arxiv.org/abs/1110.0530}
  {arXiv:1110.0530 [gr-qc]} \BibitemShut {NoStop}%
\bibitem [{\citenamefont {Sampson}\ \emph {et~al.}(2013)\citenamefont
  {Sampson}, \citenamefont {Cornish},\ and\ \citenamefont
  {Yunes}}]{Sampson:2013lpa}%
  \BibitemOpen
  \bibfield  {author} {\bibinfo {author} {\bibfnamefont {L.}~\bibnamefont
  {Sampson}}, \bibinfo {author} {\bibfnamefont {N.}~\bibnamefont {Cornish}}, \
  and\ \bibinfo {author} {\bibfnamefont {N.}~\bibnamefont {Yunes}},\ }\href
  {\doibase 10.1103/PhysRevD.87.102001} {\bibfield  {journal} {\bibinfo
  {journal} {Phys.Rev.}\ }\textbf {\bibinfo {volume} {D87}},\ \bibinfo {pages}
  {102001} (\bibinfo {year} {2013})},\ \Eprint {http://arxiv.org/abs/1303.1185}
  {arXiv:1303.1185 [gr-qc]} \BibitemShut {NoStop}%
\bibitem [{\citenamefont {Flanagan}\ and\ \citenamefont
  {Hinderer}(2008)}]{Flanagan:2007ix}%
  \BibitemOpen
  \bibfield  {author} {\bibinfo {author} {\bibfnamefont {E.~E.}\ \bibnamefont
  {Flanagan}}\ and\ \bibinfo {author} {\bibfnamefont {T.}~\bibnamefont
  {Hinderer}},\ }\href {\doibase 10.1103/PhysRevD.77.021502} {\bibfield
  {journal} {\bibinfo  {journal} {Phys.Rev.}\ }\textbf {\bibinfo {volume}
  {D77}},\ \bibinfo {pages} {021502} (\bibinfo {year} {2008})},\ \Eprint
  {http://arxiv.org/abs/0709.1915} {arXiv:0709.1915 [astro-ph]} \BibitemShut
  {NoStop}%
\bibitem [{\citenamefont {Read}\ \emph {et~al.}(2009)\citenamefont {Read},
  \citenamefont {Markakis}, \citenamefont {Shibata}, \citenamefont {Uryu},
  \citenamefont {Creighton} \emph {et~al.}}]{Read:2009yp}%
  \BibitemOpen
  \bibfield  {author} {\bibinfo {author} {\bibfnamefont {J.~S.}\ \bibnamefont
  {Read}}, \bibinfo {author} {\bibfnamefont {C.}~\bibnamefont {Markakis}},
  \bibinfo {author} {\bibfnamefont {M.}~\bibnamefont {Shibata}}, \bibinfo
  {author} {\bibfnamefont {K.}~\bibnamefont {Uryu}}, \bibinfo {author}
  {\bibfnamefont {J.~D.}\ \bibnamefont {Creighton}},  \emph {et~al.},\ }\href
  {\doibase 10.1103/PhysRevD.79.124033} {\bibfield  {journal} {\bibinfo
  {journal} {Phys.Rev.}\ }\textbf {\bibinfo {volume} {D79}},\ \bibinfo {pages}
  {124033} (\bibinfo {year} {2009})},\ \Eprint {http://arxiv.org/abs/0901.3258}
  {arXiv:0901.3258 [gr-qc]} \BibitemShut {NoStop}%
\bibitem [{\citenamefont {Lackey}\ \emph {et~al.}(2014)\citenamefont {Lackey},
  \citenamefont {Kyutoku}, \citenamefont {Shibata}, \citenamefont {Brady},\
  and\ \citenamefont {Friedman}}]{Lackey:2013axa}%
  \BibitemOpen
  \bibfield  {author} {\bibinfo {author} {\bibfnamefont {B.~D.}\ \bibnamefont
  {Lackey}}, \bibinfo {author} {\bibfnamefont {K.}~\bibnamefont {Kyutoku}},
  \bibinfo {author} {\bibfnamefont {M.}~\bibnamefont {Shibata}}, \bibinfo
  {author} {\bibfnamefont {P.~R.}\ \bibnamefont {Brady}}, \ and\ \bibinfo
  {author} {\bibfnamefont {J.~L.}\ \bibnamefont {Friedman}},\ }\href {\doibase
  10.1103/PhysRevD.89.043009} {\bibfield  {journal} {\bibinfo  {journal}
  {Phys.Rev.}\ }\textbf {\bibinfo {volume} {D89}},\ \bibinfo {pages} {043009}
  (\bibinfo {year} {2014})},\ \Eprint {http://arxiv.org/abs/1303.6298}
  {arXiv:1303.6298 [gr-qc]} \BibitemShut {NoStop}%
\bibitem [{\citenamefont {Del~Pozzo}\ \emph {et~al.}(2013)\citenamefont
  {Del~Pozzo}, \citenamefont {Li}, \citenamefont {Agathos}, \citenamefont {Van
  Den~Broeck},\ and\ \citenamefont {Vitale}}]{DelPozzo:2013ala}%
  \BibitemOpen
  \bibfield  {author} {\bibinfo {author} {\bibfnamefont {W.}~\bibnamefont
  {Del~Pozzo}}, \bibinfo {author} {\bibfnamefont {T.~G.~F.}\ \bibnamefont
  {Li}}, \bibinfo {author} {\bibfnamefont {M.}~\bibnamefont {Agathos}},
  \bibinfo {author} {\bibfnamefont {C.}~\bibnamefont {Van Den~Broeck}}, \ and\
  \bibinfo {author} {\bibfnamefont {S.}~\bibnamefont {Vitale}},\ }\href
  {\doibase 10.1103/PhysRevLett.111.071101} {\bibfield  {journal} {\bibinfo
  {journal} {Phys. Rev. Lett.}\ }\textbf {\bibinfo {volume} {111}},\ \bibinfo
  {pages} {071101} (\bibinfo {year} {2013})},\ \Eprint
  {http://arxiv.org/abs/1307.8338} {arXiv:1307.8338 [gr-qc]} \BibitemShut
  {NoStop}%
\bibitem [{\citenamefont {{Wade}}\ \emph {et~al.}(2014)\citenamefont {{Wade}},
  \citenamefont {{Creighton}}, \citenamefont {{Ochsner}}, \citenamefont
  {{Lackey}}, \citenamefont {{Farr}}, \citenamefont {{Littenberg}},\ and\
  \citenamefont {{Raymond}}}]{Wade:2014vqa}%
  \BibitemOpen
  \bibfield  {author} {\bibinfo {author} {\bibfnamefont {L.}~\bibnamefont
  {{Wade}}}, \bibinfo {author} {\bibfnamefont {J.~D.~E.}\ \bibnamefont
  {{Creighton}}}, \bibinfo {author} {\bibfnamefont {E.}~\bibnamefont
  {{Ochsner}}}, \bibinfo {author} {\bibfnamefont {B.~D.}\ \bibnamefont
  {{Lackey}}}, \bibinfo {author} {\bibfnamefont {B.~F.}\ \bibnamefont
  {{Farr}}}, \bibinfo {author} {\bibfnamefont {T.~B.}\ \bibnamefont
  {{Littenberg}}}, \ and\ \bibinfo {author} {\bibfnamefont {V.}~\bibnamefont
  {{Raymond}}},\ }\href {\doibase 10.1103/PhysRevD.89.103012} {\bibfield
  {journal} {\bibinfo  {journal} {Phys. Rev. D}\ }\textbf {\bibinfo {volume}
  {89}},\ \bibinfo {eid} {103012} (\bibinfo {year} {2014})},\ \Eprint
  {http://arxiv.org/abs/1402.5156} {arXiv:1402.5156 [gr-qc]} \BibitemShut
  {NoStop}%
\bibitem [{\citenamefont {Berti}\ \emph {et~al.}(2015)\citenamefont {Berti}
  \emph {et~al.}}]{Berti:2015itd}%
  \BibitemOpen
  \bibfield  {author} {\bibinfo {author} {\bibfnamefont {E.}~\bibnamefont
  {Berti}} \emph {et~al.},\ }\href@noop {} {\  (\bibinfo {year} {2015})},\
  \Eprint {http://arxiv.org/abs/1501.07274} {arXiv:1501.07274 [gr-qc]}
  \BibitemShut {NoStop}%
\bibitem [{\citenamefont {Belczynski}\ \emph {et~al.}(2010)\citenamefont
  {Belczynski}, \citenamefont {Bulik}, \citenamefont {Fryer}, \citenamefont
  {Ruiter}, \citenamefont {Vink} \emph {et~al.}}]{Belczynski:2009xy}%
  \BibitemOpen
  \bibfield  {author} {\bibinfo {author} {\bibfnamefont {K.}~\bibnamefont
  {Belczynski}}, \bibinfo {author} {\bibfnamefont {T.}~\bibnamefont {Bulik}},
  \bibinfo {author} {\bibfnamefont {C.~L.}\ \bibnamefont {Fryer}}, \bibinfo
  {author} {\bibfnamefont {A.}~\bibnamefont {Ruiter}}, \bibinfo {author}
  {\bibfnamefont {J.~S.}\ \bibnamefont {Vink}},  \emph {et~al.},\ }\href
  {\doibase 10.1088/0004-637X/714/2/1217} {\bibfield  {journal} {\bibinfo
  {journal} {Astrophys. J.}\ }\textbf {\bibinfo {volume} {714}},\ \bibinfo
  {pages} {1217} (\bibinfo {year} {2010})},\ \Eprint
  {http://arxiv.org/abs/arXiv:0904.2784 [astro-ph.SR]} {arXiv:0904.2784
  [astro-ph.SR]} \BibitemShut {NoStop}%
\bibitem [{\citenamefont {Belczynski}\ \emph {et~al.}(2012)\citenamefont
  {Belczynski}, \citenamefont {Wiktorowicz}, \citenamefont {Fryer},
  \citenamefont {Holz},\ and\ \citenamefont {Kalogera}}]{Belczynski:2011bn}%
  \BibitemOpen
  \bibfield  {author} {\bibinfo {author} {\bibfnamefont {K.}~\bibnamefont
  {Belczynski}}, \bibinfo {author} {\bibfnamefont {G.}~\bibnamefont
  {Wiktorowicz}}, \bibinfo {author} {\bibfnamefont {C.}~\bibnamefont {Fryer}},
  \bibinfo {author} {\bibfnamefont {D.}~\bibnamefont {Holz}}, \ and\ \bibinfo
  {author} {\bibfnamefont {V.}~\bibnamefont {Kalogera}},\ }\href {\doibase
  10.1088/0004-637X/757/1/91} {\bibfield  {journal} {\bibinfo  {journal}
  {Astrophys.J.}\ }\textbf {\bibinfo {volume} {757}},\ \bibinfo {pages} {91}
  (\bibinfo {year} {2012})},\ \Eprint {http://arxiv.org/abs/1110.1635}
  {arXiv:1110.1635 [astro-ph.GA]} \BibitemShut {NoStop}%
\bibitem [{\citenamefont {Mandel}\ \emph {et~al.}(2015)\citenamefont {Mandel},
  \citenamefont {Haster}, \citenamefont {Dominik},\ and\ \citenamefont
  {Belczynsk}}]{Mandel:2015spa}%
  \BibitemOpen
  \bibfield  {author} {\bibinfo {author} {\bibfnamefont {I.}~\bibnamefont
  {Mandel}}, \bibinfo {author} {\bibfnamefont {C.-J.}\ \bibnamefont {Haster}},
  \bibinfo {author} {\bibfnamefont {M.}~\bibnamefont {Dominik}}, \ and\
  \bibinfo {author} {\bibfnamefont {K.}~\bibnamefont {Belczynsk}},\ }\href
  {\doibase 10.1093/mnrasl/slv054} {\  (\bibinfo {year} {2015}),\
  10.1093/mnrasl/slv054},\ \Eprint {http://arxiv.org/abs/1503.03172}
  {arXiv:1503.03172 [astro-ph.HE]} \BibitemShut {NoStop}%
\bibitem [{\citenamefont {Schutz}(1986)}]{Schutz:1986gp}%
  \BibitemOpen
  \bibfield  {author} {\bibinfo {author} {\bibfnamefont {B.~F.}\ \bibnamefont
  {Schutz}},\ }\href {\doibase 10.1038/323310a0} {\bibfield  {journal}
  {\bibinfo  {journal} {Nature}\ }\textbf {\bibinfo {volume} {323}},\ \bibinfo
  {pages} {310} (\bibinfo {year} {1986})}\BibitemShut {NoStop}%
\bibitem [{\citenamefont {Holz}\ and\ \citenamefont
  {Hughes}(2005)}]{Holz:2005df}%
  \BibitemOpen
  \bibfield  {author} {\bibinfo {author} {\bibfnamefont {D.~E.}\ \bibnamefont
  {Holz}}\ and\ \bibinfo {author} {\bibfnamefont {S.~A.}\ \bibnamefont
  {Hughes}},\ }\href {\doibase 10.1086/431341} {\bibfield  {journal} {\bibinfo
  {journal} {Astrophys. J.}\ }\textbf {\bibinfo {volume} {629}},\ \bibinfo
  {pages} {15} (\bibinfo {year} {2005})},\ \Eprint
  {http://arxiv.org/abs/astro-ph/0504616} {arXiv:astro-ph/0504616 [astro-ph]}
  \BibitemShut {NoStop}%
\bibitem [{\citenamefont {Nissanke}\ \emph {et~al.}(2010)\citenamefont
  {Nissanke}, \citenamefont {Holz}, \citenamefont {Hughes}, \citenamefont
  {Dalal},\ and\ \citenamefont {Sievers}}]{Nissanke:2009kt}%
  \BibitemOpen
  \bibfield  {author} {\bibinfo {author} {\bibfnamefont {S.}~\bibnamefont
  {Nissanke}}, \bibinfo {author} {\bibfnamefont {D.~E.}\ \bibnamefont {Holz}},
  \bibinfo {author} {\bibfnamefont {S.~A.}\ \bibnamefont {Hughes}}, \bibinfo
  {author} {\bibfnamefont {N.}~\bibnamefont {Dalal}}, \ and\ \bibinfo {author}
  {\bibfnamefont {J.~L.}\ \bibnamefont {Sievers}},\ }\href {\doibase
  10.1088/0004-637X/725/1/496} {\bibfield  {journal} {\bibinfo  {journal}
  {Astrophys. J.}\ }\textbf {\bibinfo {volume} {725}},\ \bibinfo {pages} {496}
  (\bibinfo {year} {2010})},\ \Eprint {http://arxiv.org/abs/{arXiv:0904.1017
  [astro-ph.CO]}} {{arXiv:0904.1017 [astro-ph.CO]}} \BibitemShut {NoStop}%
\bibitem [{\citenamefont {Gamerman}(1997)}]{Gamerman:1997}%
  \BibitemOpen
  \bibfield  {author} {\bibinfo {author} {\bibfnamefont {D.}~\bibnamefont
  {Gamerman}},\ }\href@noop {} {\emph {\bibinfo {title} {Markov Chain Monte
  Carlo: Stocastic Simulation of Bayesian Inference}}}\ (\bibinfo  {publisher}
  {Chapman and Hall},\ \bibinfo {address} {London},\ \bibinfo {year}
  {1997})\BibitemShut {NoStop}%
\bibitem [{\citenamefont {{Skilling}}(2004)}]{Skilling}%
  \BibitemOpen
  \bibfield  {author} {\bibinfo {author} {\bibfnamefont {J.}~\bibnamefont
  {{Skilling}}},\ }in\ \href {\doibase 10.1063/1.1835238} {\emph {\bibinfo
  {booktitle} {American Institute of Physics Conference Series}}},\ \bibinfo
  {series} {American Institute of Physics Conference Series}, Vol.\ \bibinfo
  {volume} {735},\ \bibinfo {editor} {edited by\ \bibinfo {editor}
  {\bibfnamefont {R.}~\bibnamefont {{Fischer}}}, \bibinfo {editor}
  {\bibfnamefont {R.}~\bibnamefont {{Preuss}}}, \ and\ \bibinfo {editor}
  {\bibfnamefont {U.~V.}\ \bibnamefont {{Toussaint}}}}\ (\bibinfo {year}
  {2004})\ pp.\ \bibinfo {pages} {395--405}\BibitemShut {NoStop}%
\bibitem [{\citenamefont {Feroz}\ and\ \citenamefont
  {Hobson}(2007)}]{Feroz:2007kg}%
  \BibitemOpen
  \bibfield  {author} {\bibinfo {author} {\bibfnamefont {F.}~\bibnamefont
  {Feroz}}\ and\ \bibinfo {author} {\bibfnamefont {M.~P.}\ \bibnamefont
  {Hobson}},\ }\href@noop {} {\  (\bibinfo {year} {2007})},\ \Eprint
  {http://arxiv.org/abs/arXiv:0704.3704} {arXiv:arXiv:0704.3704 [astro-ph]}
  \BibitemShut {NoStop}%
\bibitem [{\citenamefont {{Metzger}}\ \emph {et~al.}(2010)\citenamefont
  {{Metzger}}, \citenamefont {{Mart{\'{\i}}nez-Pinedo}}, \citenamefont
  {{Darbha}}, \citenamefont {{Quataert}}, \citenamefont {{Arcones}},
  \citenamefont {{Kasen}}, \citenamefont {{Thomas}}, \citenamefont {{Nugent}},
  \citenamefont {{Panov}},\ and\ \citenamefont {{Zinner}}}]{Metzger:2010}%
  \BibitemOpen
  \bibfield  {author} {\bibinfo {author} {\bibfnamefont {B.~D.}\ \bibnamefont
  {{Metzger}}}, \bibinfo {author} {\bibfnamefont {G.}~\bibnamefont
  {{Mart{\'{\i}}nez-Pinedo}}}, \bibinfo {author} {\bibfnamefont
  {S.}~\bibnamefont {{Darbha}}}, \bibinfo {author} {\bibfnamefont
  {E.}~\bibnamefont {{Quataert}}}, \bibinfo {author} {\bibfnamefont
  {A.}~\bibnamefont {{Arcones}}}, \bibinfo {author} {\bibfnamefont
  {D.}~\bibnamefont {{Kasen}}}, \bibinfo {author} {\bibfnamefont
  {R.}~\bibnamefont {{Thomas}}}, \bibinfo {author} {\bibfnamefont
  {P.}~\bibnamefont {{Nugent}}}, \bibinfo {author} {\bibfnamefont {I.~V.}\
  \bibnamefont {{Panov}}}, \ and\ \bibinfo {author} {\bibfnamefont {N.~T.}\
  \bibnamefont {{Zinner}}},\ }\href {\doibase 10.1111/j.1365-2966.2010.16864.x}
  {\bibfield  {journal} {\bibinfo  {journal} {Mon. Not. Roy. Astron. Soc.}\
  }\textbf {\bibinfo {volume} {406}},\ \bibinfo {pages} {2650} (\bibinfo {year}
  {2010})},\ \Eprint {http://arxiv.org/abs/1001.5029} {arXiv:1001.5029
  [astro-ph.HE]} \BibitemShut {NoStop}%
\bibitem [{\citenamefont {Pankow}\ \emph {et~al.}(2015)\citenamefont {Pankow},
  \citenamefont {Brady}, \citenamefont {Ochsner},\ and\ \citenamefont
  {O'Shaughnessy}}]{Pankow:2015cra}%
  \BibitemOpen
  \bibfield  {author} {\bibinfo {author} {\bibfnamefont {C.}~\bibnamefont
  {Pankow}}, \bibinfo {author} {\bibfnamefont {P.}~\bibnamefont {Brady}},
  \bibinfo {author} {\bibfnamefont {E.}~\bibnamefont {Ochsner}}, \ and\
  \bibinfo {author} {\bibfnamefont {R.}~\bibnamefont {O'Shaughnessy}},\ }\href
  {\doibase 10.1103/PhysRevD.92.023002} {\bibfield  {journal} {\bibinfo
  {journal} {Phys. Rev.}\ }\textbf {\bibinfo {volume} {D92}},\ \bibinfo {pages}
  {023002} (\bibinfo {year} {2015})},\ \Eprint
  {http://arxiv.org/abs/1502.04370} {arXiv:1502.04370 [gr-qc]} \BibitemShut
  {NoStop}%
\bibitem [{\citenamefont {{Haster}}\ \emph {et~al.}(2015)\citenamefont
  {{Haster}}, \citenamefont {{Mandel}},\ and\ \citenamefont
  {{Farr}}}]{Haster:2015}%
  \BibitemOpen
  \bibfield  {author} {\bibinfo {author} {\bibfnamefont {C.-J.}\ \bibnamefont
  {{Haster}}}, \bibinfo {author} {\bibfnamefont {I.}~\bibnamefont {{Mandel}}},
  \ and\ \bibinfo {author} {\bibfnamefont {W.~M.}\ \bibnamefont {{Farr}}},\
  }\href@noop {} {\bibfield  {journal} {\bibinfo  {journal} {ArXiv e-prints}\ }
  (\bibinfo {year} {2015})},\ \Eprint {http://arxiv.org/abs/1502.05407}
  {arXiv:1502.05407 [astro-ph.IM]} \BibitemShut {NoStop}%
\bibitem [{\citenamefont {Farr}(2015)}]{Farr:2015}%
  \BibitemOpen
  \bibfield  {author} {\bibinfo {author} {\bibfnamefont {B.}~\bibnamefont
  {Farr}},\ }\href@noop {} {\enquote {\bibinfo {title} {Kombine},}\ } (\bibinfo
  {year} {2015}),\ \bibinfo {note} {\url{http://home.uchicago.edu/~farr/kombine/}}\BibitemShut {NoStop}%
\bibitem [{\citenamefont {{Farr}}\ \emph {et~al.}(2014)\citenamefont {{Farr}},
  \citenamefont {{Kalogera}},\ and\ \citenamefont {{Luijten}}}]{Farr:2014}%
  \BibitemOpen
  \bibfield  {author} {\bibinfo {author} {\bibfnamefont {B.}~\bibnamefont
  {{Farr}}}, \bibinfo {author} {\bibfnamefont {V.}~\bibnamefont {{Kalogera}}},
  \ and\ \bibinfo {author} {\bibfnamefont {E.}~\bibnamefont {{Luijten}}},\
  }\href {\doibase 10.1103/PhysRevD.90.024014} {\bibfield  {journal} {\bibinfo
  {journal} {\prd}\ }\textbf {\bibinfo {volume} {90}},\ \bibinfo {eid} {024014}
  (\bibinfo {year} {2014})},\ \Eprint {http://arxiv.org/abs/1309.7709}
  {arXiv:1309.7709 [astro-ph.IM]} \BibitemShut {NoStop}%
\bibitem [{\citenamefont {Singer}\ and\ \citenamefont
  {Price}(2015)}]{Singer:2015ema}%
  \BibitemOpen
  \bibfield  {author} {\bibinfo {author} {\bibfnamefont {L.~P.}\ \bibnamefont
  {Singer}}\ and\ \bibinfo {author} {\bibfnamefont {L.~R.}\ \bibnamefont
  {Price}},\ }\href@noop {} {\  (\bibinfo {year} {2015})},\ \Eprint
  {http://arxiv.org/abs/1508.03634} {arXiv:1508.03634 [gr-qc]} \BibitemShut
  {NoStop}%
\bibitem [{\citenamefont {Singer}\ \emph {et~al.}(2014)\citenamefont {Singer}
  \emph {et~al.}}]{Singer:2014qca}%
  \BibitemOpen
  \bibfield  {author} {\bibinfo {author} {\bibfnamefont {L.~P.}\ \bibnamefont
  {Singer}} \emph {et~al.},\ }\href {\doibase 10.1088/0004-637X/795/2/105}
  {\bibfield  {journal} {\bibinfo  {journal} {Astrophys. J.}\ }\textbf
  {\bibinfo {volume} {795}},\ \bibinfo {pages} {105} (\bibinfo {year}
  {2014})},\ \Eprint {http://arxiv.org/abs/1404.5623} {arXiv:1404.5623
  [astro-ph.HE]} \BibitemShut {NoStop}%
\bibitem [{\citenamefont {Berry}\ \emph {et~al.}(2015)\citenamefont {Berry}
  \emph {et~al.}}]{Berry:2014jja}%
  \BibitemOpen
  \bibfield  {author} {\bibinfo {author} {\bibfnamefont {C.~P.~L.}\
  \bibnamefont {Berry}} \emph {et~al.},\ }\href {\doibase
  10.1088/0004-637X/804/2/114} {\bibfield  {journal} {\bibinfo  {journal}
  {Astrophys. J.}\ }\textbf {\bibinfo {volume} {804}},\ \bibinfo {pages} {114}
  (\bibinfo {year} {2015})},\ \Eprint {http://arxiv.org/abs/1411.6934}
  {arXiv:1411.6934 [astro-ph.HE]} \BibitemShut {NoStop}%
\bibitem [{\citenamefont {Farr}\ \emph {et~al.}(2015)\citenamefont {Farr} \emph
  {et~al.}}]{Farr:2015lna}%
  \BibitemOpen
  \bibfield  {author} {\bibinfo {author} {\bibfnamefont {B.}~\bibnamefont
  {Farr}} \emph {et~al.},\ }\href@noop {} {\  (\bibinfo {year} {2015})},\
  \Eprint {http://arxiv.org/abs/1508.05336} {arXiv:1508.05336 [astro-ph.HE]}
  \BibitemShut {NoStop}%
\bibitem [{\citenamefont {{Fragos}}\ and\ \citenamefont
  {{McClintock}}(2015)}]{Fragos:2015}%
  \BibitemOpen
  \bibfield  {author} {\bibinfo {author} {\bibfnamefont {T.}~\bibnamefont
  {{Fragos}}}\ and\ \bibinfo {author} {\bibfnamefont {J.~E.}\ \bibnamefont
  {{McClintock}}},\ }\href {\doibase 10.1088/0004-637X/800/1/17} {\bibfield
  {journal} {\bibinfo  {journal} {\apj}\ }\textbf {\bibinfo {volume} {800}},\
  \bibinfo {eid} {17} (\bibinfo {year} {2015})},\ \Eprint
  {http://arxiv.org/abs/1408.2661} {arXiv:1408.2661 [astro-ph.HE]} \BibitemShut
  {NoStop}%
\bibitem [{\citenamefont {Miller}\ \emph {et~al.}(2015)\citenamefont {Miller},
  \citenamefont {O'Shaughnessy}, \citenamefont {Littenberg},\ and\
  \citenamefont {Farr}}]{Miller:2015sga}%
  \BibitemOpen
  \bibfield  {author} {\bibinfo {author} {\bibfnamefont {B.}~\bibnamefont
  {Miller}}, \bibinfo {author} {\bibfnamefont {R.}~\bibnamefont
  {O'Shaughnessy}}, \bibinfo {author} {\bibfnamefont {T.~B.}\ \bibnamefont
  {Littenberg}}, \ and\ \bibinfo {author} {\bibfnamefont {B.}~\bibnamefont
  {Farr}},\ }\href@noop {} {\  (\bibinfo {year} {2015})},\ \Eprint
  {http://arxiv.org/abs/1506.06032} {arXiv:1506.06032 [gr-qc]} \BibitemShut
  {NoStop}%
\bibitem [{\citenamefont {Farr}\ \emph {et~al.}(2014)\citenamefont {Farr},
  \citenamefont {Ochsner}, \citenamefont {Farr},\ and\ \citenamefont
  {O'Shaughnessy}}]{Farr:2014qka}%
  \BibitemOpen
  \bibfield  {author} {\bibinfo {author} {\bibfnamefont {B.}~\bibnamefont
  {Farr}}, \bibinfo {author} {\bibfnamefont {E.}~\bibnamefont {Ochsner}},
  \bibinfo {author} {\bibfnamefont {W.~M.}\ \bibnamefont {Farr}}, \ and\
  \bibinfo {author} {\bibfnamefont {R.}~\bibnamefont {O'Shaughnessy}},\ }\href
  {\doibase 10.1103/PhysRevD.90.024018} {\bibfield  {journal} {\bibinfo
  {journal} {Phys. Rev.}\ }\textbf {\bibinfo {volume} {D90}},\ \bibinfo {pages}
  {024018} (\bibinfo {year} {2014})},\ \Eprint {http://arxiv.org/abs/1404.7070}
  {arXiv:1404.7070 [gr-qc]} \BibitemShut {NoStop}%
\bibitem [{\citenamefont {Littenberg}\ \emph {et~al.}(2015)\citenamefont
  {Littenberg}, \citenamefont {Farr}, \citenamefont {Coughlin}, \citenamefont
  {Kalogera},\ and\ \citenamefont {Holz}}]{Littenberg:2015tpa}%
  \BibitemOpen
  \bibfield  {author} {\bibinfo {author} {\bibfnamefont {T.~B.}\ \bibnamefont
  {Littenberg}}, \bibinfo {author} {\bibfnamefont {B.}~\bibnamefont {Farr}},
  \bibinfo {author} {\bibfnamefont {S.}~\bibnamefont {Coughlin}}, \bibinfo
  {author} {\bibfnamefont {V.}~\bibnamefont {Kalogera}}, \ and\ \bibinfo
  {author} {\bibfnamefont {D.~E.}\ \bibnamefont {Holz}},\ }\href {\doibase
  10.1088/2041-8205/807/2/L24} {\bibfield  {journal} {\bibinfo  {journal}
  {Astrophys. J.}\ }\textbf {\bibinfo {volume} {807}},\ \bibinfo {pages} {L24}
  (\bibinfo {year} {2015})},\ \Eprint {http://arxiv.org/abs/1503.03179}
  {arXiv:1503.03179 [astro-ph.HE]} \BibitemShut {NoStop}%
\bibitem [{\citenamefont {Aasi}\ \emph {et~al.}(2013)\citenamefont {Aasi} \emph
  {et~al.}}]{ObservingScenarios}%
  \BibitemOpen
  \bibfield  {author} {\bibinfo {author} {\bibfnamefont {J.}~\bibnamefont
  {Aasi}} \emph {et~al.} (\bibinfo {collaboration} {LIGO Scientific
  Collaboration, Virgo Collaboration}),\ }\href@noop {} {\  (\bibinfo {year}
  {2013})},\ \Eprint {http://arxiv.org/abs/1304.0670} {arXiv:1304.0670 [gr-qc]}
  \BibitemShut {NoStop}%
\bibitem [{LAL()}]{LAL}%
  \BibitemOpen
  \href@noop {} {\enquote {\bibinfo {title} {{LSC Algorithm Library suite}},}\
  }\bibinfo {howpublished}
  {\url{https://www.lsc-group.phys.uwm.edu/daswg/projects/lalsuite.html}}\BibitemShut
  {NoStop}%
\bibitem [{\citenamefont {Veitch}\ \emph {et~al.}(2015)\citenamefont {Veitch},
  \citenamefont {Raymond}, \citenamefont {Farr}, \citenamefont {Farr},
  \citenamefont {Graff} \emph {et~al.}}]{LALInference}%
  \BibitemOpen
  \bibfield  {author} {\bibinfo {author} {\bibfnamefont {J.}~\bibnamefont
  {Veitch}}, \bibinfo {author} {\bibfnamefont {V.}~\bibnamefont {Raymond}},
  \bibinfo {author} {\bibfnamefont {B.}~\bibnamefont {Farr}}, \bibinfo {author}
  {\bibfnamefont {W.}~\bibnamefont {Farr}}, \bibinfo {author} {\bibfnamefont
  {P.}~\bibnamefont {Graff}},  \emph {et~al.},\ }\href {\doibase
  10.1103/PhysRevD.91.042003} {\bibfield  {journal} {\bibinfo  {journal}
  {Phys.Rev.}\ }\textbf {\bibinfo {volume} {D91}},\ \bibinfo {pages} {042003}
  (\bibinfo {year} {2015})},\ \Eprint {http://arxiv.org/abs/1409.7215}
  {arXiv:1409.7215 [gr-qc]} \BibitemShut {NoStop}%
\bibitem [{\citenamefont {Farr}\ \emph {et~al.}(2011)\citenamefont {Farr},
  \citenamefont {Sravan}, \citenamefont {Cantrell}, \citenamefont {Kreidberg},
  \citenamefont {Bailyn} \emph {et~al.}}]{Farr:2010tu}%
  \BibitemOpen
  \bibfield  {author} {\bibinfo {author} {\bibfnamefont {W.~M.}\ \bibnamefont
  {Farr}}, \bibinfo {author} {\bibfnamefont {N.}~\bibnamefont {Sravan}},
  \bibinfo {author} {\bibfnamefont {A.}~\bibnamefont {Cantrell}}, \bibinfo
  {author} {\bibfnamefont {L.}~\bibnamefont {Kreidberg}}, \bibinfo {author}
  {\bibfnamefont {C.~D.}\ \bibnamefont {Bailyn}},  \emph {et~al.},\ }\href
  {\doibase 10.1088/0004-637X/741/2/103} {\bibfield  {journal} {\bibinfo
  {journal} {Astrophys.J.}\ }\textbf {\bibinfo {volume} {741}},\ \bibinfo
  {pages} {103} (\bibinfo {year} {2011})},\ \Eprint
  {http://arxiv.org/abs/1011.1459} {arXiv:1011.1459 [astro-ph.GA]} \BibitemShut
  {NoStop}%
\bibitem [{\citenamefont {{\"O}zel}\ \emph {et~al.}(2010)\citenamefont
  {{\"O}zel}, \citenamefont {Psaltis}, \citenamefont {Narayan},\ and\
  \citenamefont {McClintock}}]{Ozel:2010su}%
  \BibitemOpen
  \bibfield  {author} {\bibinfo {author} {\bibfnamefont {F.}~\bibnamefont
  {{\"O}zel}}, \bibinfo {author} {\bibfnamefont {D.}~\bibnamefont {Psaltis}},
  \bibinfo {author} {\bibfnamefont {R.}~\bibnamefont {Narayan}}, \ and\
  \bibinfo {author} {\bibfnamefont {J.~E.}\ \bibnamefont {McClintock}},\ }\href
  {\doibase 10.1088/0004-637X/725/2/1918} {\bibfield  {journal} {\bibinfo
  {journal} {Astrophys.J.}\ }\textbf {\bibinfo {volume} {725}},\ \bibinfo
  {pages} {1918} (\bibinfo {year} {2010})},\ \Eprint
  {http://arxiv.org/abs/1006.2834} {arXiv:1006.2834 [astro-ph.GA]} \BibitemShut
  {NoStop}%
\bibitem [{\citenamefont {Kreidberg}\ \emph {et~al.}(2012)\citenamefont
  {Kreidberg}, \citenamefont {Bailyn}, \citenamefont {Farr},\ and\
  \citenamefont {Kalogera}}]{Kreidberg:2012ud}%
  \BibitemOpen
  \bibfield  {author} {\bibinfo {author} {\bibfnamefont {L.}~\bibnamefont
  {Kreidberg}}, \bibinfo {author} {\bibfnamefont {C.~D.}\ \bibnamefont
  {Bailyn}}, \bibinfo {author} {\bibfnamefont {W.~M.}\ \bibnamefont {Farr}}, \
  and\ \bibinfo {author} {\bibfnamefont {V.}~\bibnamefont {Kalogera}},\ }\href
  {\doibase 10.1088/0004-637X/757/1/36} {\bibfield  {journal} {\bibinfo
  {journal} {Astrophys.J.}\ }\textbf {\bibinfo {volume} {757}},\ \bibinfo
  {pages} {36} (\bibinfo {year} {2012})},\ \Eprint
  {http://arxiv.org/abs/1205.1805} {arXiv:1205.1805 [astro-ph.HE]} \BibitemShut
  {NoStop}%
\bibitem [{\citenamefont {Nissanke}\ \emph {et~al.}(2013)\citenamefont
  {Nissanke}, \citenamefont {Kasliwal},\ and\ \citenamefont
  {Georgieva}}]{Nissanke:2012dj}%
  \BibitemOpen
  \bibfield  {author} {\bibinfo {author} {\bibfnamefont {S.}~\bibnamefont
  {Nissanke}}, \bibinfo {author} {\bibfnamefont {M.}~\bibnamefont {Kasliwal}},
  \ and\ \bibinfo {author} {\bibfnamefont {A.}~\bibnamefont {Georgieva}},\
  }\href {\doibase 10.1088/0004-637X/767/2/124} {\bibfield  {journal} {\bibinfo
   {journal} {Astrophys. J.}\ }\textbf {\bibinfo {volume} {767}},\ \bibinfo
  {pages} {124} (\bibinfo {year} {2013})},\ \Eprint
  {http://arxiv.org/abs/1210.6362} {arXiv:1210.6362 [astro-ph.HE]} \BibitemShut
  {NoStop}%
\bibitem [{\citenamefont {Hanna}\ \emph {et~al.}(2014)\citenamefont {Hanna},
  \citenamefont {Mandel},\ and\ \citenamefont {Vousden}}]{Hanna:2013yda}%
  \BibitemOpen
  \bibfield  {author} {\bibinfo {author} {\bibfnamefont {C.}~\bibnamefont
  {Hanna}}, \bibinfo {author} {\bibfnamefont {I.}~\bibnamefont {Mandel}}, \
  and\ \bibinfo {author} {\bibfnamefont {W.}~\bibnamefont {Vousden}},\ }\href
  {\doibase 10.1088/0004-637X/784/1/8} {\bibfield  {journal} {\bibinfo
  {journal} {Astrophys. J.}\ }\textbf {\bibinfo {volume} {784}},\ \bibinfo
  {pages} {8} (\bibinfo {year} {2014})},\ \Eprint
  {http://arxiv.org/abs/1312.2077} {arXiv:1312.2077 [astro-ph.HE]} \BibitemShut
  {NoStop}%
\bibitem [{\citenamefont {Fan}\ \emph {et~al.}(2014)\citenamefont {Fan},
  \citenamefont {Messenger},\ and\ \citenamefont {Heng}}]{Fan:2014kka}%
  \BibitemOpen
  \bibfield  {author} {\bibinfo {author} {\bibfnamefont {X.}~\bibnamefont
  {Fan}}, \bibinfo {author} {\bibfnamefont {C.}~\bibnamefont {Messenger}}, \
  and\ \bibinfo {author} {\bibfnamefont {I.~S.}\ \bibnamefont {Heng}},\ }\href
  {\doibase 10.1088/0004-637X/795/1/43} {\bibfield  {journal} {\bibinfo
  {journal} {Astrophys. J.}\ }\textbf {\bibinfo {volume} {795}},\ \bibinfo
  {pages} {43} (\bibinfo {year} {2014})},\ \Eprint
  {http://arxiv.org/abs/1406.1544} {arXiv:1406.1544 [astro-ph.HE]} \BibitemShut
  {NoStop}%
\bibitem [{\citenamefont {Cutler}\ and\ \citenamefont
  {Flanagan}(1994)}]{Cutler:1994ys}%
  \BibitemOpen
  \bibfield  {author} {\bibinfo {author} {\bibfnamefont {C.}~\bibnamefont
  {Cutler}}\ and\ \bibinfo {author} {\bibfnamefont {E.~E.}\ \bibnamefont
  {Flanagan}},\ }\href {\doibase 10.1103/PhysRevD.49.2658} {\bibfield
  {journal} {\bibinfo  {journal} {Phys. Rev. D}\ }\textbf {\bibinfo {volume}
  {49}},\ \bibinfo {pages} {2658} (\bibinfo {year} {1994})},\ \Eprint
  {http://arxiv.org/abs/{arXiv:gr-qc/9402014}} {{arXiv:gr-qc/9402014}}
  \BibitemShut {NoStop}%
\bibitem [{\citenamefont {Rodriguez}\ \emph {et~al.}(2014)\citenamefont
  {Rodriguez}, \citenamefont {Farr}, \citenamefont {Raymond}, \citenamefont
  {Farr}, \citenamefont {Littenberg} \emph {et~al.}}]{Rodriguez:2013oaa}%
  \BibitemOpen
  \bibfield  {author} {\bibinfo {author} {\bibfnamefont {C.~L.}\ \bibnamefont
  {Rodriguez}}, \bibinfo {author} {\bibfnamefont {B.}~\bibnamefont {Farr}},
  \bibinfo {author} {\bibfnamefont {V.}~\bibnamefont {Raymond}}, \bibinfo
  {author} {\bibfnamefont {W.~M.}\ \bibnamefont {Farr}}, \bibinfo {author}
  {\bibfnamefont {T.~B.}\ \bibnamefont {Littenberg}},  \emph {et~al.},\ }\href
  {\doibase 10.1088/0004-637X/784/2/119} {\bibfield  {journal} {\bibinfo
  {journal} {Astrophys.J.}\ }\textbf {\bibinfo {volume} {784}},\ \bibinfo
  {pages} {119} (\bibinfo {year} {2014})},\ \Eprint
  {http://arxiv.org/abs/1309.3273} {arXiv:1309.3273 [astro-ph.HE]} \BibitemShut
  {NoStop}%
\bibitem [{\citenamefont {Field}\ \emph {et~al.}(2014)\citenamefont {Field},
  \citenamefont {Galley}, \citenamefont {Hesthaven}, \citenamefont {Kaye},\
  and\ \citenamefont {Tiglio}}]{Field:2013cfa}%
  \BibitemOpen
  \bibfield  {author} {\bibinfo {author} {\bibfnamefont {S.~E.}\ \bibnamefont
  {Field}}, \bibinfo {author} {\bibfnamefont {C.~R.}\ \bibnamefont {Galley}},
  \bibinfo {author} {\bibfnamefont {J.~S.}\ \bibnamefont {Hesthaven}}, \bibinfo
  {author} {\bibfnamefont {J.}~\bibnamefont {Kaye}}, \ and\ \bibinfo {author}
  {\bibfnamefont {M.}~\bibnamefont {Tiglio}},\ }\href {\doibase
  10.1103/PhysRevX.4.031006} {\bibfield  {journal} {\bibinfo  {journal} {Phys.
  Rev.}\ }\textbf {\bibinfo {volume} {X4}},\ \bibinfo {pages} {031006}
  (\bibinfo {year} {2014})},\ \Eprint {http://arxiv.org/abs/1308.3565}
  {arXiv:1308.3565 [gr-qc]} \BibitemShut {NoStop}%
\bibitem [{\citenamefont {Lundgren}\ and\ \citenamefont
  {O'Shaughnessy}(2014)}]{Lundgren:2013jla}%
  \BibitemOpen
  \bibfield  {author} {\bibinfo {author} {\bibfnamefont {A.}~\bibnamefont
  {Lundgren}}\ and\ \bibinfo {author} {\bibfnamefont {R.}~\bibnamefont
  {O'Shaughnessy}},\ }\href {\doibase 10.1103/PhysRevD.89.044021} {\bibfield
  {journal} {\bibinfo  {journal} {Phys.Rev.}\ }\textbf {\bibinfo {volume}
  {D89}},\ \bibinfo {pages} {044021} (\bibinfo {year} {2014})},\ \Eprint
  {http://arxiv.org/abs/1304.3332} {arXiv:1304.3332 [gr-qc]} \BibitemShut
  {NoStop}%
\bibitem [{\citenamefont {Chatziioannou}\ \emph {et~al.}(2014)\citenamefont
  {Chatziioannou}, \citenamefont {Cornish}, \citenamefont {Klein},\ and\
  \citenamefont {Yunes}}]{Chatziioannou:2014bma}%
  \BibitemOpen
  \bibfield  {author} {\bibinfo {author} {\bibfnamefont {K.}~\bibnamefont
  {Chatziioannou}}, \bibinfo {author} {\bibfnamefont {N.}~\bibnamefont
  {Cornish}}, \bibinfo {author} {\bibfnamefont {A.}~\bibnamefont {Klein}}, \
  and\ \bibinfo {author} {\bibfnamefont {N.}~\bibnamefont {Yunes}},\ }\href
  {\doibase 10.1103/PhysRevD.89.104023} {\bibfield  {journal} {\bibinfo
  {journal} {Phys. Rev.}\ }\textbf {\bibinfo {volume} {D89}},\ \bibinfo {pages}
  {104023} (\bibinfo {year} {2014})},\ \Eprint {http://arxiv.org/abs/1404.3180}
  {arXiv:1404.3180 [gr-qc]} \BibitemShut {NoStop}%
\bibitem [{\citenamefont {Klein}\ \emph {et~al.}(2014)\citenamefont {Klein},
  \citenamefont {Cornish},\ and\ \citenamefont {Yunes}}]{Klein:2014bua}%
  \BibitemOpen
  \bibfield  {author} {\bibinfo {author} {\bibfnamefont {A.}~\bibnamefont
  {Klein}}, \bibinfo {author} {\bibfnamefont {N.}~\bibnamefont {Cornish}}, \
  and\ \bibinfo {author} {\bibfnamefont {N.}~\bibnamefont {Yunes}},\ }\href
  {\doibase 10.1103/PhysRevD.90.124029} {\bibfield  {journal} {\bibinfo
  {journal} {Phys. Rev.}\ }\textbf {\bibinfo {volume} {D90}},\ \bibinfo {pages}
  {124029} (\bibinfo {year} {2014})},\ \Eprint {http://arxiv.org/abs/1408.5158}
  {arXiv:1408.5158 [gr-qc]} \BibitemShut {NoStop}%
\end{thebibliography}

%

\end{document}